\documentclass[twocolumn]{aastex62}

\hypersetup{linkcolor=red,citecolor=blue,filecolor=cyan,urlcolor=magenta}
\usepackage{natbib}
\usepackage{amsthm}
\usepackage{amssymb}
\usepackage{amsmath}
\usepackage{bm}

\received{\today}
\revised{, 2018}
\accepted{, 2018}
\submitjournal{ApJ}

\shorttitle{Firehose instability}
\shortauthors{Shaaban et al.}

\begin{document}

\title{The Interplay of the Solar Wind Core and Suprathermal Electrons: A Quasilinear Approach 
for Firehose Instability}


\author[0000-0003-0465-598X]{S.M.Shaaban}
\affil{Centre for Mathematical Plasma Astrophysics, KU Leuven, Celestijnenlaan 200B, B-3001 Leuven, Belgium}
\affiliation{Theoretical Physics Research Group, Physics Dept., Faculty of Science, Mansoura University, 35516, Mansoura, Egypt}

\author[0000-0002-8508-5466]{M. Lazar}
\affiliation{Centre for Mathematical Plasma Astrophysics, KU Leuven, Celestijnenlaan 200B, B-3001 Leuven, Belgium}
\affiliation{Institut f\"ur Theoretische Physik, Lehrstuhl IV: Weltraum- und Astrophysik, Ruhr-Universit\"at Bochum, D-44780 Bochum, Germany}

\author[0000-0001-8134-3790]{P.H.Yoon}
\affiliation{Institute for Physical Science and Technology, University of Maryland, College Park, MD 20742, USA}
\affiliation{Korea Astronomy and Space Science Institute, Daejeon 34055, Korea}
\affiliation{School of Space Research, Kyung Hee University, Yongin, Gyeonggi 17104, Korea}

\author[0000-0002-1743-0651]{S. Poedts}
\affiliation{Centre for Mathematical Plasma Astrophysics, KU Leuven, Celestijnenlaan 200B, B-3001 Leuven, Belgium}

\begin{abstract}

In the solar wind an equipartition of kinetic energy densities can be easily established 
between thermal and suprathermal electrons and the instability conditions are markedly 
altered by the interplay of these two populations. The new thresholds derived here for 
the periodic branch of firehose instability shape the limits of temperature anisotropy 
reported by the observations for both electron populations. This instability 
constraint is particularly important for the suprathermal electrons which, by comparison 
to thermal populations, are even less controlled by the particle-particle collisions. An 
extended quasilinear approach of this instability confirms predictions from linear theory 
and unveil the mutual effects of thermal and suprathermal electrons in the relaxation of 
their temperature anisotropies and the saturation of growing fluctuations.

\end{abstract}


\section{Introduction} \label{sec:1}

In collision-poor plasmas in space, e.g., the solar wind and planetary environments, 
the decay of free energy from large scales is mediated by the instabilities and the observed
enhanced fluctuations \citep{Stverak2008, Gary2015, Gershman2017}. For instance, the beam-plasma instabilities should play a 
major role in the dissipation of solar plasma outflows, e.g., from coronal holes 
or during coronal mass ejections (CMEs) \citep{Ganse2012, Jian2014}, while an increase of temperature 
anisotropy $T_\parallel > T_\perp$ (where $\parallel, \perp$ refer to directions relative to the 
interplanetary magnetic field), as predicted by the CGL invariants \citep{Chew1956} at large 
heliocentric distances (where particle-particle collisions are inefficient), is expected to 
be constrained by the firehose instability \citep{Stverak2008, Lazar2017a}. In a quiet solar wind, e.g., during slow 
winds, the observations seem to confirm a potential role of this instability, and that is, 
surprisingly, for the quasi-thermal (bi-Maxwellian) populations whose large deviations from 
isotropy appear to be well shaped by the instability thresholds predicted by the kinetic 
theory \citep{Kasper2006, Hellinger2006, Stverak2008}. However, an important amount of 
kinetic (free) energy is transported by the suprathermal particle populations, which are 
ubiquitous in space plasmas and are well described by the Kappa power-laws 
\citep{Vasyliunas1968, Pierrard2010}. We need therefore to take into account the effects 
of these populations for a realistic description of firehose instability in the 
solar wind context.

Recent attempts to characterize these effects are either limited to a linear analysis \citep{Lazar2009, Lazar2015, Lazar2017b, Vinas2017}, or simply altered by an idealized (bi-)Maxwellian description of suprathermal populations in a quasilinear analysis. Thus, quasilinear approaches have been successfully proposed for both the  whistler and firehose instabilities driven either by a single bi-Maxwellian population of electrons \citep{Yoon2012, Yoon2017a, Sarfraz2017}, or by dual electrons with a core-halo structure \citep{Sarfraz2016, Sarfraz2018} minimizing, however, the effects of suprathermals by assuming the halo bi-Maxwellian distributed.  Here we present 
a quasilinear approach of firehose instability driven by the anisotropic electrons, i.e., 
$A \equiv T_\parallel/T_\perp < 1$, adopting an advanced dual model for the electron velocity 
distributions, as indicated by the observations in the solar wind \citep{Stverak2008, 
Maksimovic2005, Pierrard2016}. This model combines two main components, a thermal 
bi-Maxwellian core at low energies, and a suprathermal bi-Kappa halo which enhances the 
high-energy tails of the distributions. An additional strahl can be detected (mainly 
during fast winds) streaming anti-sunwards along the magnetic field, but at large enough distances 
from the Sun, e.g., beyond 1~AU, the strahl diminishes considerably (probably scattered by the 
self-generated instabilities) and a dual core-halo composition remains fairly dominant 
\citep{Maksimovic2005}. 

The instability develops from the interplay between the core (subscript $c$)
and halo (subscript $h$) electrons, and the highest growing modes (with the lowest thresholds) are 
therefore expected to arise from a cumulative effect when both populations exhibit similar
anisotropies $A_{c,h} < 1$. 
We assume a homogenous plasma dominated by the electrons and protons, and neglect the effects of 
any other minor species that can be present in the solar wind. In section \ref{sec:2} we first introduce
the velocity distribution functions, a dual core-halo model for the electrons while heavier protons 
are assumed Maxwellian and isotropic, and then build the linear and quasilinear formalisms used to 
describe the dispersion and stability of firehose solutions. Derived numerically these solutions 
are discussed for several representative cases in section~\ref{sec:3}. We restrict to firehose 
modes propagating parallel to the magnetic field, especially because of the complexity of a 
quasilinear theory which becomes less feasible for an arbitrary propagation. By contrast to previous
studies, here we adopt a new normalization for the wave parameters in order to avoid any artificial 
coupling between the core and halo electrons, which may alter the quasilinear relaxation under 
the influence of increasing firehose fluctuations.
In section \ref{sec:4} we summarize the results and provide the main conclusions of this study.

\section{Modeling based on observations}\label{sec:2}

The velocity distribution functions (VDFs) of plasma particles and, in general, parametrization used for 
our magnetized plasma approach are characteristic to the solar wind at different heliocentric 
distances, but may also be relevant for more particular environments like planetary magnetospheres. 

\subsection{Solar wind electrons}

The in-situ measurements collected from different missions, e.g., Ulysses, Helios 1, and Cluster 
II, unveil electron distributions in the slow winds ($V_{SW}\leqslant500$ km s$^{-1}$) with a 
dual structure combining a thermal dense core (subscript $c$), and a dilute suprathermal 
halo (subscript $c$) \citep{Maksimovic2005, Stverak2008}
\begin{equation}\label{e1}
f_e\left(v_\parallel,v_\perp \right)=\eta_c~f_c\left(v_\parallel,v_\perp \right)+ \eta_h
~f_h\left(v_\parallel,v_\perp \right).  
\end{equation}
Here, $\eta_h=n_h/n_0$ and $\eta_c=1-\eta_h$ are relative densities of the halo and core, respectively, 
and $n_0$ is the total electron number density. The core population is well fitted by a 
bi-Maxwellian distribution \citep{Stverak2008}
\begin{align}
 \label{e2}
f_{c}\left( v_{\parallel },v_{\perp }\right) =&\frac{1}{\pi
^{3/2}\alpha_{\perp, c }^{2} ~ \alpha_{\parallel, c }}\exp \left(
-\frac{v_{\parallel }^{2}} {\alpha_{\parallel, c}^{2}}-\frac{v_{\perp
}^{2}}{\alpha_{\perp, c}^{2}}\right),   
\end{align}
with thermal velocities $\alpha_{ \parallel, \perp, c} \equiv\alpha_{\parallel, \perp, c}(t)$ (varying in time $t$ in our quasilinear approach) defined by the corresponding temperature 
components (as the second-order moments of the distribution) 
\label{e3}
\begin{subequations}
\begin{align}
T_{\parallel, c}&=\frac{m_e}{k_B}\int d\textbf{v} v_\parallel^2
f_c(v_\parallel, v_\perp)=\frac{m_e ~\alpha_{\parallel, c}^2}{2 k_B},\\
T_{\perp, c}&=\frac{m_e}{2 k_B}\int d\textbf{v} v_\perp^2
f_c(v_\parallel, v_\perp)=\frac{m_e ~\alpha_{\perp, c}^2}{2 k_B}.
\end{align}
\end{subequations}
The halo component is described by an anisotropic bi-Kappa distribution function
\begin{eqnarray} \label{e4}
f_{h}\left( v_{\parallel },v_{\perp }\right)=&&\frac{1}{\pi ^{3/2} \alpha_{\perp, h}^{2}~ \alpha_{\parallel, h}}
\frac{\Gamma\left( \kappa +1\right)}{\Gamma \left( \kappa -1/2\right)}\nonumber\\
&&\times \left[ 1+\frac{v_{\parallel }^{2}}{\kappa~\alpha_{ \parallel, h }^{2}}
+\frac{v_{\perp }^{2}}{\kappa~\alpha_{ \perp, h }^{2}}\right] ^{-\kappa-1},
\end{eqnarray}
with parameters $\alpha_{ \parallel, \perp, h} \equiv\alpha_{\parallel, \perp, h}(t)$ (varying in time $t$ in our quasilinear approach) defined by the components of the anisotropic temperature
\begin{align}\label{e5}
T_{\parallel, h}=&\frac{2 \kappa}{2 \kappa-3}\frac{m_e }{2 k_B}\alpha_{\parallel, h}^2,~T_{\perp, h}=&\frac{2 \kappa}{2 \kappa-3}\frac{m_e}{2 k_B}\alpha_{\perp, h}^2.
\end{align}
\begin{figure}[t]
\centering
\includegraphics[scale=0.5]{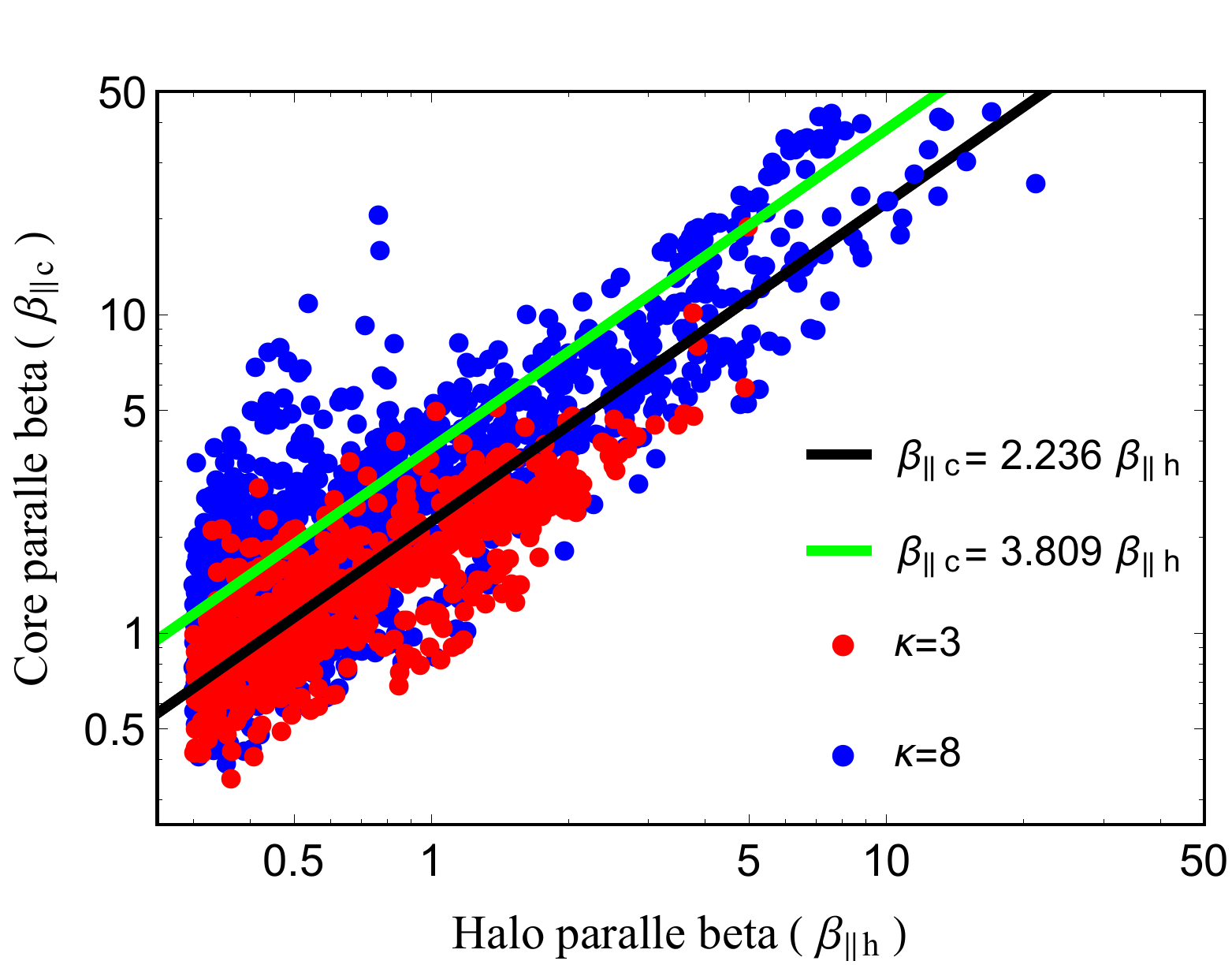}
\caption{Plasma beta values measured in the solar wind for the core and halo electrons with the 
apparent linear correlations, for $\kappa=3$ (red dots and black fit) and 
$\kappa = 8$ (blue dots and green fit).}\label{f1}
\end{figure}

Such a plasma system we parametrize by inspiring from the observations reported by 
\cite{Stverak2008}, from roughly 120000 events detected in the ecliptic at different 
distances ($0.3-3.95$ AU) from the Sun. We chose only the slow wind data ($V_{SW}\leqslant500$ 
km s$^{-1}$), for which the density of the strahl population is faint enough to not affect 
the instability conditions. These data have been used in a series of recent studies 
to characterize the electron core and halo populations \citep{Pierrard2016, Lazar2017b}. 
For slow wind conditions, the suprathermal halo tails are fitted by a Kappa with $2.5 <\kappa
< 9.5$. Comparing to the core population, the halo is hotter ($T_h > T_c$) but less dense, 
with a number density $\eta_h \leqslant 0.10$ \citep{Maksimovic2005}. Both the core and 
halo populations may exhibit temperature anisotropy, i.e., $A=T_\perp/T_\parallel \neq 1$, 
which can be at the origin of kinetic instabilities, such as whistler (if $A >1$) or firehose 
instability (if $A < 1$). Their anisotropies show a prevalent tendency for
a direct (linear) correlation, like $A_c=C A_h $, with $C\simeq 1$ \citep{Pierrard2016}. In the 
present study we are interested for the conditions most favorable to firehose instability. 
For the interplay of the core and suprathermal halo these conditions are those associated with 
similar sub-unitary anisotropies $A_c=A_h<1$. We use the observations to evaluate the 
relative densities $\eta_h$ and $\eta_c$ and the ratio of parallel plasma betas $\beta_c/
\beta_h$ (where $\beta_{\parallel}\equiv 8\pi n k_B T_{\parallel}/B^2$), corresponding to two 
different representative values of $\kappa-$index. Figure \ref{f1} displays a density histogram 
for $\beta_c$ vs. $\beta_h$ with a color bar counting the number of events. For a low $\kappa=3$ 
($2.5\leqslant\kappa \leqslant3.5$) we find relevant $\beta_c/\beta_h \simeq 2.236$, and for a 
high $\kappa=8$ ($7.5\leqslant\kappa \leqslant8.5$) we consider $\beta_c/\beta_h\simeq 3.809$. 
Contrary to other recent studies which assume $\beta_c < \beta_h$, e.g., \citet[and refs. therein] 
{Sarfraz2018}, here we consider $\beta_c > \beta_h$ as suggested by the observations. A similar 
linear correlation between plasma beta parameters $\beta_c$ and $\beta_h$ is found in \cite{Lazar2015} 
using a different set of data from Ulysses during the slow winds. In Section \ref{sec:3} 
we show how important this ratio is for contrasting the core and halo parallel betas for the 
instability conditions.  

\subsection{Theory of dispersion and stability}

For a collisionless and homogeneous plasma, the general, linear dispersion relation of the 
electromagnetic modes propagating parallel to the background magnetic field (${\bf B}_0$), 
i.e., ${\bf k} \times {\bf B}_0 = 0$) reads
\begin{eqnarray}\label{e6}
 {k^2c^2 \over \omega^2} = 1  +
\sum_{a=p,c,h} {\omega_{p, a}^2 \over \omega^2} \int d\bm{v} {v_\perp\over 2 
\left(\omega - k v_{\parallel} \pm \Omega_a\right)} \nonumber&&\\
\left((\omega - k v_{\parallel}) {\partial f_{a} \over \partial v_{\perp}} + 
k v_{\perp} {\partial f_{a} \over \partial v_{\parallel}} \right),
\end{eqnarray}
where $k$ is the wave number, $c$ is the speed of light, $\omega_{p, \alpha}= \sqrt{4\pi n_a 
e^2/m_a}$ and $\Omega_a=e B_0/m_a c$ are, respectively, the non-relativistic plasma frequency 
and the gyro-frequency of species $a$, and $\pm$ denotes the right-handed (RH) and left-handed 
(LH) circular polarization, respectively. 

For the LH electron firehose unstable modes in our plasma system the instantaneous dispersion relation (\ref{e6}) reduces to

\begin{eqnarray} \label{e10}
\tilde{k}^2=&&A_p-1+\left(\frac{A_p~\tilde{\omega} -\left(A_p-1\right)}{\tilde{k} \sqrt{\beta_p}}\right)Z_p\left(\frac{\tilde{\omega}-1}{\tilde{k} \sqrt{\beta_p}}\right)\nonumber\\
&&+\eta_c~\mu \left[A_c-1+\left(\frac{A_c~\tilde{\omega} +\left(A_c-1\right)\mu}{\tilde{k} \sqrt{\mu~\beta_c/\eta_c}}\right)\right.\nonumber\\
&&\left.\times Z_c\left(\frac{\tilde{\omega}+\mu}{\tilde{k} \sqrt{\mu~\beta_c/\eta_c}}\right)\right]+\eta_h~\mu \left[A_h-1\right.\\
&&\left.+ \left(\frac{A_h~\tilde{\omega} +\left(A_h-1\right)\mu}{\tilde{k} \sqrt{\mu~\chi_h~\beta_h/\eta_h}}\right)Z_{h}\left(\frac{\tilde{\omega}+\mu}{\tilde{k} \sqrt{\mu~\chi_h~\beta_h/\eta_h}}\right)\right]\nonumber
\end{eqnarray} 
where $\tilde{k}=kc/\omega_{p,p}$ is the normalized wave-number, $\tilde{\omega}=\omega/\Omega_p$ 
is the normalized wave frequency, $\mu=m_p/m_e$ is the proton--electron mass 
contrast, $A_a=~T_{\perp,a}/T_{\parallel,a}\equiv~\beta_{\perp,a}/\beta_{\parallel,a}$ and $\beta_{\parallel,\perp , a}=8\pi n_a k_B T_{\parallel \perp , a}/B_0^2$ are, respectively, the 
temperature anisotropy, and plasma beta parameters for protons (subscript "$a=p$"), electron 
core (subscript "$a=c$"), and electron halo (subscript "$a=h$") populations, 
$\chi_h=(\kappa-1.5)/\kappa$,  $\eta_h=n_h/n_0$, $\eta_c=1-\eta_h$ are the halo and 
the core density contrast, respectively,
\begin{equation}  \label{e13}
Z_{a}\left( \xi _{a}^{\pm }\right) =\frac{1}{\sqrt{\pi}}\int_{-\infty
}^{\infty }\frac{\exp \left( -x^{2}\right) }{x-\xi _{a}^{\pm }}dt,\
\ \Im \left( \xi _{a}^{\pm }\right) >0, 
\end{equation}
is the plasma dispersion function \citep{Fried1961} and 
\begin{eqnarray} \label{e14}
     Z_h\left( \xi_{h}^{+}\right) =&&\frac{1}{\pi ^{1/2}\kappa^{1/2}}\frac{\Gamma \left( \kappa \right) }{\Gamma \left(\kappa -1/2\right) }\nonumber\\
     &&\int_{-\infty }^{\infty }\frac{\left(1+x^{2}/\kappa \right) ^{-\kappa}}{x-\xi_{h}^{+}}dx,\  \Im \left(\xi _{h}^{\pm }\right) >0.
\end{eqnarray}
is the generalized (Kappa) dispersion function \citep{Lazar2008}.

The time evolution of the velocity distributions are described by the particle kinetic equation 
in the diffusion approximation
\begin{eqnarray} \label{e7}
\frac{\partial f_a}{\partial t}&&=\frac{i e^2}{4m_a^2 c^2~ v_\perp}\int_{-\infty}^{\infty} 
\frac{dk}{k}\left[ \left(\omega^\ast-k v_\parallel\right)\frac{\partial}{\partial v_\perp}+ 
k v_\perp\frac{\partial}{\partial v_\parallel}\right]\nonumber\\
&&\times~\frac{ v_\perp \delta B^2(k, \omega)}{\omega-kv_\parallel-\Omega_a}\left[ 
\left(\omega-k v_\parallel\right)\frac{\partial f_a}{\partial v_\perp}+ k v_\perp
\frac{\partial f_a}{\partial v_\parallel}\right], 
\end{eqnarray}
where the energy density of the fluctuations $\delta B^2(k)$ is described by the wave equation 
\begin{equation} \label{e9}
\frac{\partial~\delta B^2(k)}{\partial t}=2 \gamma_k \delta B^2(k),
\end{equation}
with growth rate $\gamma_k$ of the EFH instability. Eq.~\eqref{e7} is used to derive 
perpendicular and parallel velocity moments for protons, and electron core and halo populations,
as follows
\begin{subequations}\label{e8}
\begin{align}
\frac{dT_{\perp p}}{dt}&=-\frac{e^2}{2m_p c^2}
\int_{-\infty}^{\infty}\frac{dk}{k^2}\langle~ \delta B^2(k)~\rangle\\
&\times\left\lbrace\left(2 A_p-1\right)\gamma_k+\text{Im} \frac{2i\gamma-\Omega_p}{k\alpha_{\parallel p}}~ F^{-}\left(A_p, \Omega_p, Z_{p}\right)\right\rbrace\nonumber\\
\frac{dT_{\parallel p}}{dt}&=\frac{e^2}{m_p c^2}
\int_{-\infty}^{\infty}\frac{dk}{k^2}\langle~ \delta B^2(k)~\rangle\\
&\times \left\lbrace A_p~\gamma_k+\text{Im} \frac{\omega-\Omega_p}{k\alpha_{\parallel p}}~ F^{-}\left(A_p, \Omega_p, Z_{p}\right)\right\rbrace\nonumber\\
\frac{dT_{\perp c}}{dt}&=-\frac{e^2}{2m_e c^2}
\int_{-\infty}^{\infty}\frac{dk}{k^2}\langle~ \delta B^2(k)~\rangle\\
&\times\left\lbrace\left(2 A_c-1\right)\gamma_k+\text{Im} \frac{2i\gamma+\Omega_e}{k\alpha_{\parallel c}}~ F^{+}\left(A_c, \Omega_e, Z_{c}\right)\right\rbrace\nonumber\\
\frac{dT_{\parallel c}}{dt}&=\frac{e^2}{m_e c^2}
\int_{-\infty}^{\infty}\frac{dk}{k^2}\langle~ \delta B^2(k)~\rangle \\
&\times\left\lbrace A_c~\gamma_k+\text{Im} \frac{\omega+\Omega_e}{k\alpha_{\parallel c}}~ F^{+}\left(A_c, \Omega_e, Z_{c}\right)\right\rbrace\nonumber\\
\frac{dT_{\perp h}}{dt}&=-\frac{e^2}{2m_e c^2}
\int_{-\infty}^{\infty}\frac{dk}{k^2}\langle~ \delta B^2(k)~\rangle\\
&\times\left\lbrace\left(2 A_h-1\right)\gamma_k+\text{Im} \frac{2i\gamma+\Omega_e}{k\alpha_{\parallel h}}~ F^{+}\left(A_h, \Omega_e, Z_{h}\right)\right\rbrace\nonumber\\
\frac{dT_{\parallel h}}{dt}&=\frac{e^2}{m_e c^2}
\int_{-\infty}^{\infty}\frac{dk}{k^2}\langle~ \delta B^2(k)~\rangle\\
&\times \left\lbrace A_h~\gamma_k+\text{Im} \frac{\omega+\Omega_h}{k\alpha_{\parallel h}}~ F^{+}\left(A_h, \Omega_e, Z_{h}\right)\right\rbrace\nonumber
\end{align}
\end{subequations}
with
\begin{eqnarray}
F^{\pm}(A_a, \Omega_a, Z_{a} )=\left[ A_a~\omega\pm\Omega_a\left(A_a-1\right)\right]Z_{a}\left(\frac{\omega\pm\Omega_a}{k\alpha_{\parallel a}}\right).\nonumber
\end{eqnarray}
In terms of the normalized quantities these equations can be rewritten, respectively, as
\begin{subequations}\label{e11}
\begin{align}
\frac{d\beta_{\perp p}}{d\tau}=&-\int\frac{d\tilde{k}}{\tilde{k}^2} W(\tilde{k})\left\lbrace\left(2 A_p-1\right)\tilde{\gamma}+\text{Im} \frac{2i\tilde{\gamma}-1}{\tilde{k}\sqrt{\beta_{\parallel p}}}~\right.\nonumber\\
&\left.\times\left[ A_p~\tilde{\omega}-\left(A_a-1\right)\right]Z_{p}\left(\frac{\tilde{\omega}-1}{\tilde{k}\sqrt{\beta_{\parallel p}}}\right)\right\rbrace\\
\frac{d\beta_{\parallel p}}{d\tau}=&2\int\frac{d\tilde{k}}{\tilde{k}^2} W(\tilde{k})\left\lbrace  A_p~\tilde{\gamma}+\text{Im} \frac{\tilde{\omega}-1}{\tilde{k}\sqrt{\beta_{\parallel p}}}~\right.\nonumber\\
&\left.\times\left[ A_p~\tilde{\omega}-\left(A_a-1\right)\right]Z_{p}\left(\frac{\tilde{\omega}-1}{\tilde{k}\sqrt{\beta_{\parallel p}}}\right)\right\rbrace\\
\frac{d\beta_{\perp c}}{d\tau}=&-\eta_c\int\frac{d\tilde{k}}{\tilde{k}^2} W(\tilde{k})\left\lbrace \mu \left(2 A_c-1\right)\tilde{\gamma}+\text{Im} \frac{2i\tilde{\gamma}+\mu}{\tilde{k}\sqrt{\beta_{\parallel c}/\eta_c}}~~\right.\nonumber\\
&\left.\times\tilde{F}(A_c, Z_{c}, \beta_c, \eta_c )\right\rbrace\\
\frac{d\beta_{\parallel c}}{d\tau}=&2~\eta_c\int\frac{d\tilde{k}}{\tilde{k}^2} W(\tilde{k})\left\lbrace \mu~A_c~\tilde{\gamma}+\text{Im} \frac{\tilde{\omega}+\mu}{\tilde{k}\sqrt{\beta_{\parallel c}/\eta_c}}~~\right.\nonumber\\
&\left.\times\tilde{F}(A_c, Z_{c} , \beta_c, \eta_c )\right\rbrace\\
\frac{d\beta_{\perp h}}{d\tau}=&-\eta_h\int\frac{d\tilde{k}}{\tilde{k}^2} W(\tilde{k})\left\lbrace \mu \left(2 A_h-1\right)\tilde{\gamma}+\text{Im} \frac{2i\tilde{\gamma}+\mu}{\tilde{k}\sqrt{\beta_{\parallel h}/\eta_h}}~\right.\nonumber\\
&\left.\times\tilde{F}(A_h, Z_{h}, \chi_h, \beta_h, \eta_h )\right\rbrace\\
\frac{d\beta_{\parallel h}}{d\tau}=&2~\eta_h\int\frac{d\tilde{k}}{\tilde{k}^2} W(\tilde{k})\left\lbrace \mu~A_h~\tilde{\gamma}+\text{Im} \frac{\tilde{\omega}+\mu}{\tilde{k}\sqrt{\beta_{\parallel h}/\eta_h}}~\right.\nonumber\\
&\left.\times\tilde{F}(A_h, Z_{h}, \chi_h, \beta_h, \eta_h )\right\rbrace
\end{align}
\end{subequations}
with
\begin{eqnarray}
\tilde{F}(A_a, Z_{a}, \chi_a,\beta_a, \eta_a )=&&\sqrt{\mu}\left[ A_a~\tilde{\omega}+\left(A_a-1\right)\mu\right]\nonumber\\
&&\times Z_{a}\left(\frac{\tilde{\omega}+\mu}{\tilde{k}\sqrt{\mu~\chi_a~\beta_{\parallel~a}/\eta_a }}\right),\nonumber
\end{eqnarray}
and 
\begin{eqnarray}\label{e12}
\frac{\partial~W(\tilde{k})}{\partial \tau}=2~\tilde{\gamma}~ W(\tilde{k}).
\end{eqnarray}
where $W(\tilde{k})=\delta B^2(\tilde{k})/B_0^2$ is the wave 
energy density, $\tau=~\Omega_p~t$, and $\chi_c=1$.

For the normalization, we keep distinction between the plasma frequencies of the core 
(subscript $c$) and halo (subscript $h$) electrons, i.e., $\omega_{p,a}=\sqrt{4\pi n_a e^2/m_e}$ 
is defined in terms of the number density $n_a$, which is expected to play an important role
mainly triggering the effects of these components on instabilities. Normalization to a total 
plasma frequency $\omega_{p,e}= \sqrt{4\pi n_o e^2/m_e} $, where the total number density $n_0 = n_c + n_h$, invoked 
in similar studies, e.g., \citet[and refs. therein]{Sarfraz2018}, may introduce an artificial 
coupling between the core and halo electrons, and therefore alter their quasilinear relaxation 
under the effect of firehose fluctuations. 

%
\section{Numerical results} \label{sec:3}

This section presents the results of our numerical linear and quasilinear analyses of 
EFH modes for two distinct sets of plasma parameters, which we name case~1 and case~2. 
These two cases have chosen to correspond to high and low values of the power-index, e.g. 
$\kappa=8$ in case~1, and $\kappa=3$ in case 2, see parametrizations \eqref{e15} and 
\eqref{e16} below. As already motivated in Section \ref{sec:2}, this parametrization is 
suggested by the in-situ measurements of the solar wind electrons in a large interval 
($0.3-3.95$~AU) of heliocentric distances \citep{Maksimovic2005, Stverak2008, Pierrard2016, 
Lazar2017b}.

\begin{itemize}
\item{Case 1. }
 \begin{eqnarray}\label{e15}
&&\kappa=8,~ \eta_h=0.056, W(k)=10^{-6} \nonumber\\
&&A_{c,h}(0)=\frac{T_{\perp, c,h}(0)}{T_{\parallel, c,h}(0)}=\frac{\beta_{\perp, c,h}(0)}{\beta_{\parallel, c, h}(0)}=0.1,\nonumber\\
&&\beta_c(0)=5,~ 7,~ 10, \text{and}~\beta_h(0)=\beta_c(0)/3.809,\nonumber\\
&&A_p(0)=1.0,~\beta_p(0)=\beta_c(0)/3.
\end{eqnarray}
\item{Case 2. }
\begin{eqnarray} \label{e16}
&&\kappa=3,~\eta_h=0.037, W(k)=10^{-6} \nonumber\\
&&A_{c,h}(0)=\frac{T_{\perp, c,h}(0)}{T_{\parallel, c, h}(0)}=\frac{\beta_{\perp, c, h}(0)}{\beta_{\parallel, c, h}(0)}=0.1,~0.2,~0.3,\nonumber\\
&&\beta_c(0)=5,~\beta_h(0)=\beta_c(0)/2.236,\nonumber\\
&&A_p(0)=1.0,~\beta_p(0)=\beta_c(0)/3.
\end{eqnarray}
\end{itemize}

\begin{figure}[t]
\centering
\includegraphics[scale=0.53, trim={3.8cm 4.5cm 3cm 2.6cm},clip]{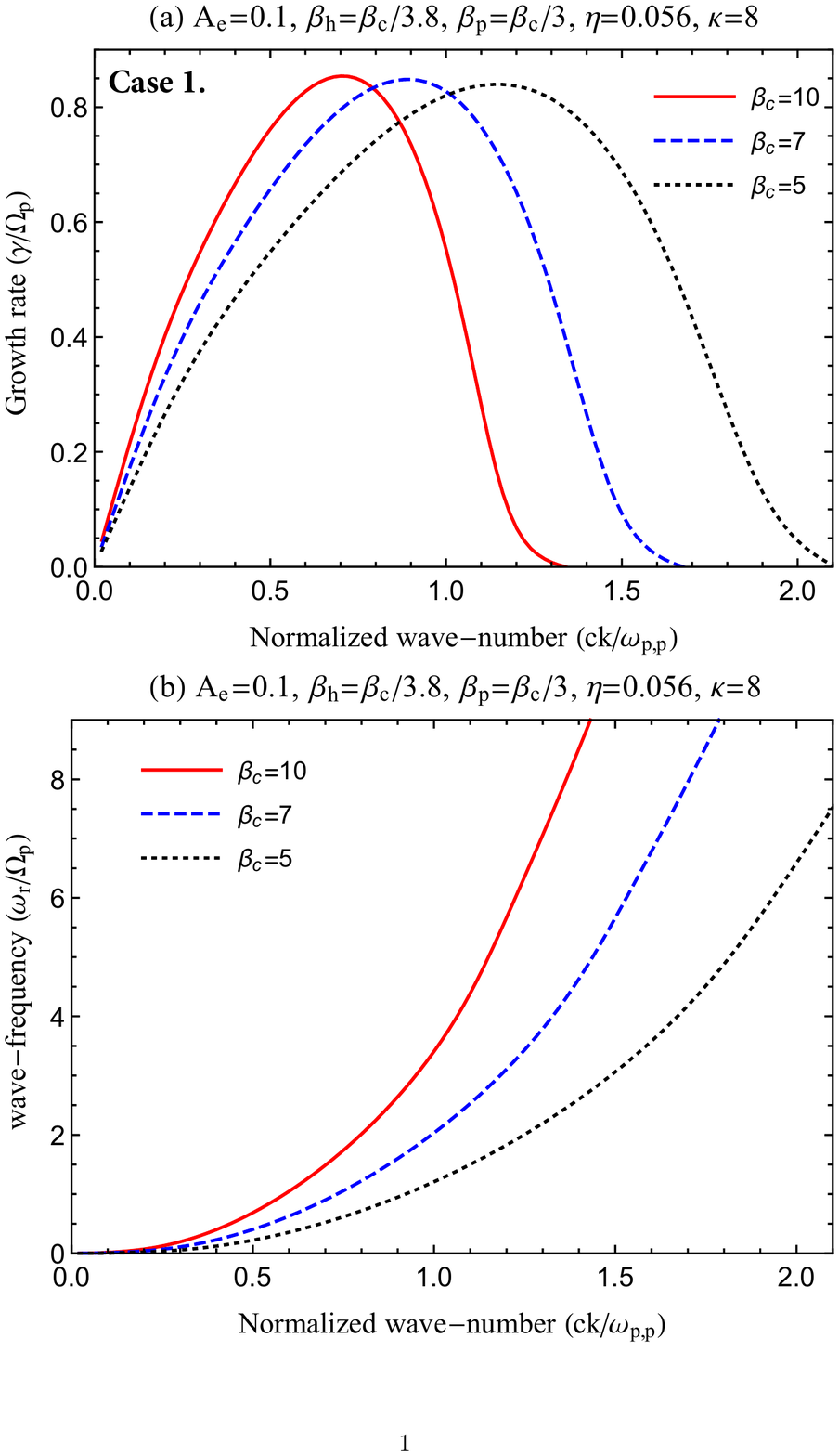}
\caption{Variation of growth rates (top) and wave-frequencies (bottom) with plasma beta.}\label{f2}
\end{figure}

\begin{figure}[t]
\centering
\includegraphics[scale=0.52, trim={3.8cm 4.5cm 3cm 2.6cm},clip]{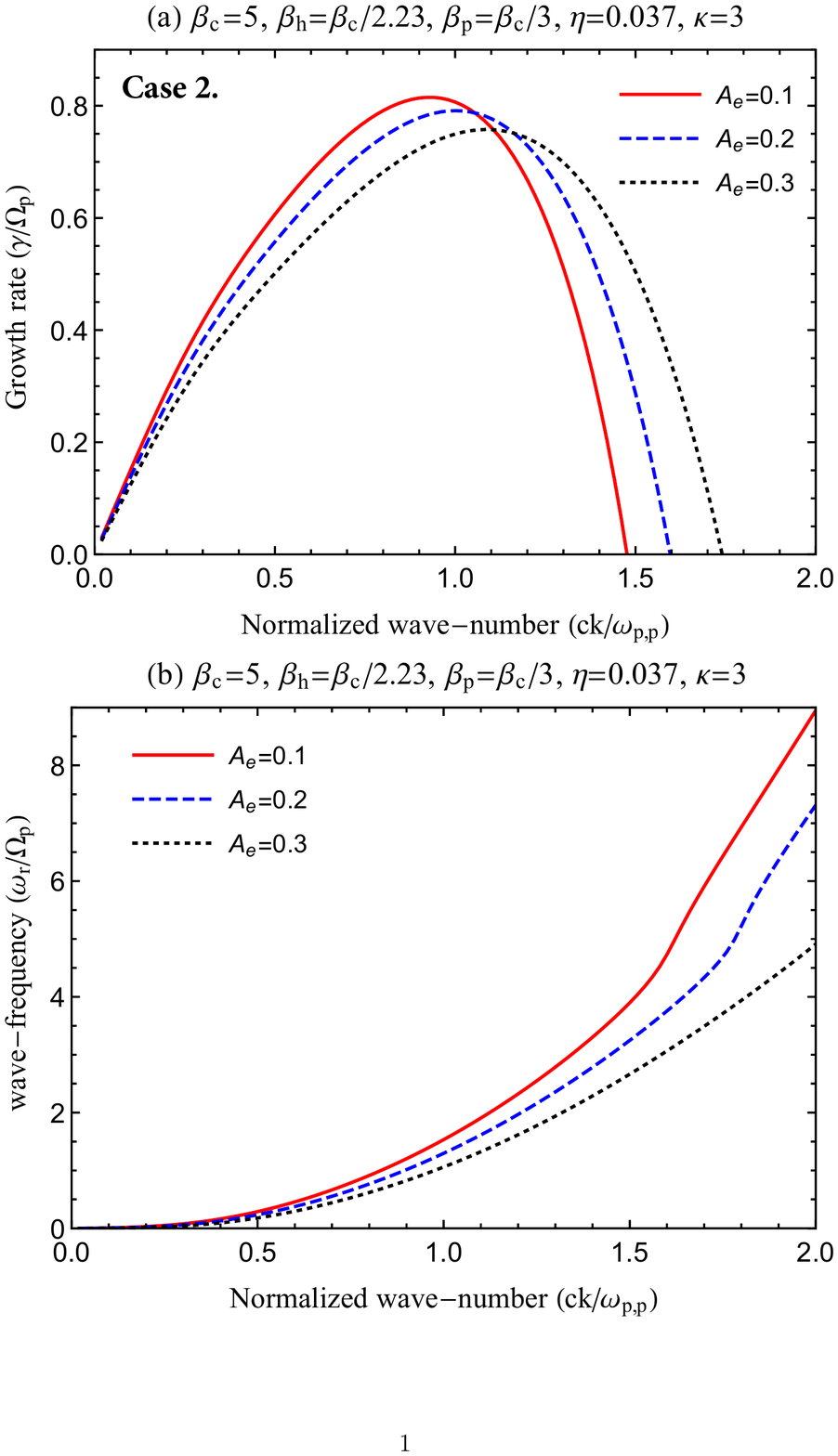}
\caption{Variation of growth rates (top) and wave-frequencies (bottom) with temperature anisotropy.}\label{f2b}
\end{figure}
%
\begin{figure}[t]
\centering
\includegraphics[scale=0.5, trim={2.65cm 3.6cm 1.8cm 2.6cm},clip]{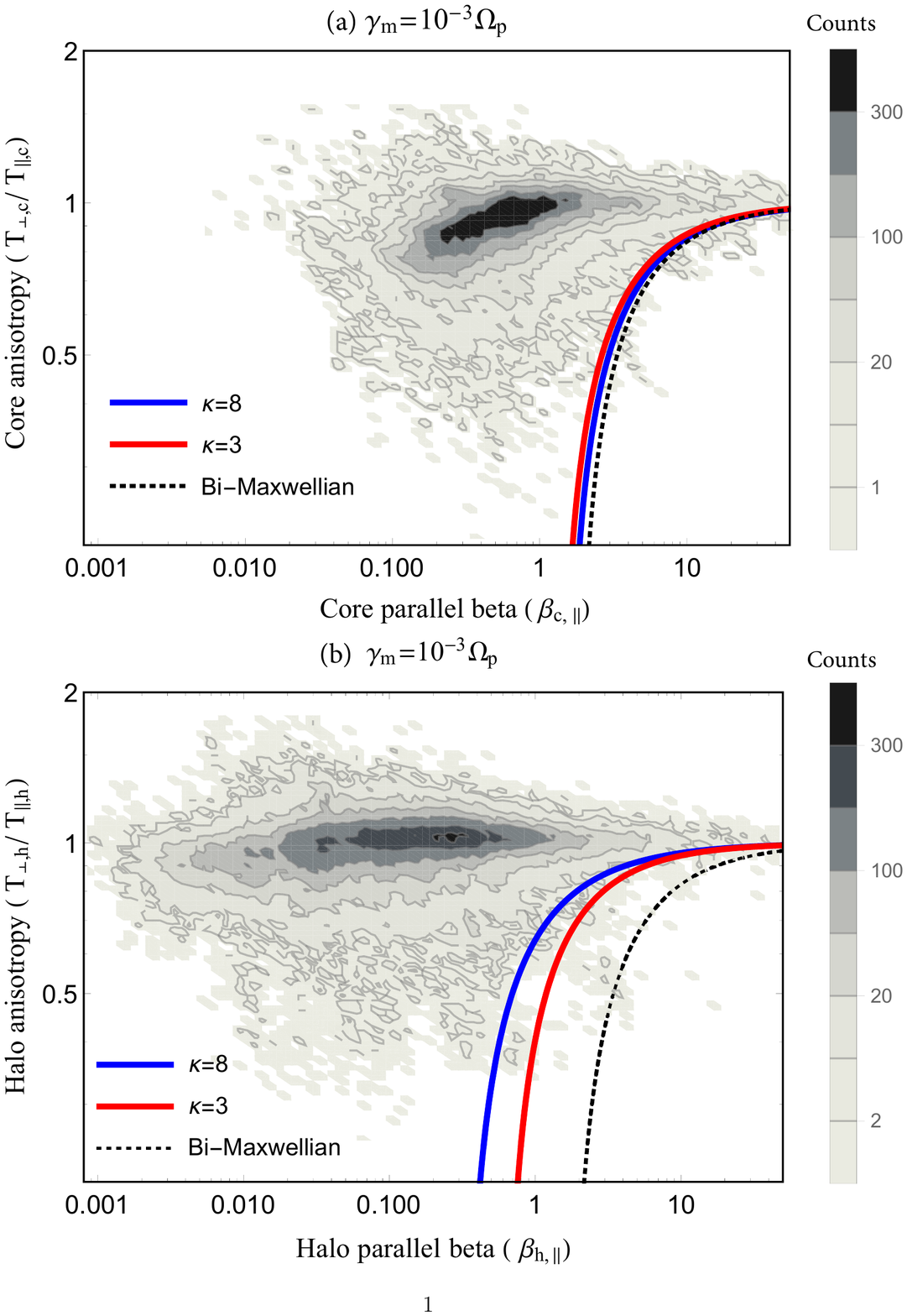}
\caption{A comparison between theoretical anisotropy thresholds and the slow solar wind observational 
data for the core (panel a), and halo (panel b) anisotropies. }\label{f3}
\end{figure}
%

\subsection{Linear analysis}\label{sec:3.1}

For a linear analysis we use the dispersion relation (\ref{e10}). Figure~\ref{f2} and \ref{f2b} illustrate unstable EFH solutions for case 1 and case 2, respectively. Case~1 ($\kappa=8$, 
$A_{c,h}=0.1$), allows us to compare growth rates for different core plasma betas 
$\beta_{\parallel, c}=5, 7, 10$ (implying different halo plasma betas, respectively, 
$\beta_{\parallel, h}=1.312,~ 1.84,~ 2.63$). Increasing the plasma beta may slightly enhance the 
the maximum growth rates (peaks), but markedly diminishes the range of unstable wavenumbers.
These counter-balancing effects are also observed for case 2 (not shown here). 

In Figure~\ref{f2b} we study the effect of the electron anisotropies $A_{c,h}(0)=0.1, 0.2, 0.3$ 
considering the plasma parameters in case 2 ($\kappa=3$), as for case 1 the observed variations
are similar. The maximum (peaking) growth rates increase as the electron anisotropy increases,
but the range of unstable wave-numbers decreases. The EFH instability becomes more operative at 
lower wave numbers by increasing the plasma beta parameters and/or electron anisotropies. The 
corresponding wave frequencies are increasing with increasing plasma betas and/or 
the electron anisotropy, see panels (b) in Figures~\ref{f2} and \ref{f2b}.  

We can already highlight the importance of these results and motivate the present study by  
a comparative analysis between theoretical predictions from linear theory and the observations in 
the solar wind. We use the same set of slow wind ($V_{sw}\leqslant~500~ $km~s$^{-1}$) data mentioned in 
Section~\ref{sec:2}, to compare the observed anisotropies with theoretical thresholds of the instability, see
Figure\ \ref{f3}. Data are displayed in Figure~\ref{f3} using a histogram plot counting for the number 
of events with color logarithmic scale. Thresholds of EFH unstable solutions are derived for a 
maximum growth rate $\gamma_m=10^{-3} \Omega_p$ and are fitted to an inverse power-law \citep{Lazar2018}
\begin{equation}\label{e17}
A_{a}=1-\frac{s}{\beta_{\parallel,a}^\alpha}
\end{equation}
for the core ($a=c$, panel a) and the halo ($a=h$, panel b), with fitting parameters $s$ and 
$\alpha$ given in Table~\ref{t1}. 

The new instability thresholds from the interplay of the core and halo population (red for 
$\kappa=3$ and blue $\kappa=8$) are contrasted with those obtained for a single bi-Maxwellian 
component (dotted black). We should observe that these new thresholds  markedly decrease 
approaching and shaping much better the limits of the halo anisotropy (panel b). The lower 
threshold is obtained in this case for a higher $\kappa=8$ (blue, case 1), under the influence of 
a more dense halo (comparing to case 2). We can also admit minor (but still visible) changes of 
the instability thresholds constraining the core anisotropy (panel a). The instability driven 
by the core populations is stimulated by the anisotropic halo, but lower thresholds are obtained
for a more dense core (red, case 2). 

\begin{figure*}[t!]
\centering
\includegraphics[scale=1.1, trim={2.58cm 12.3cm 2.cm 2.6cm},clip]{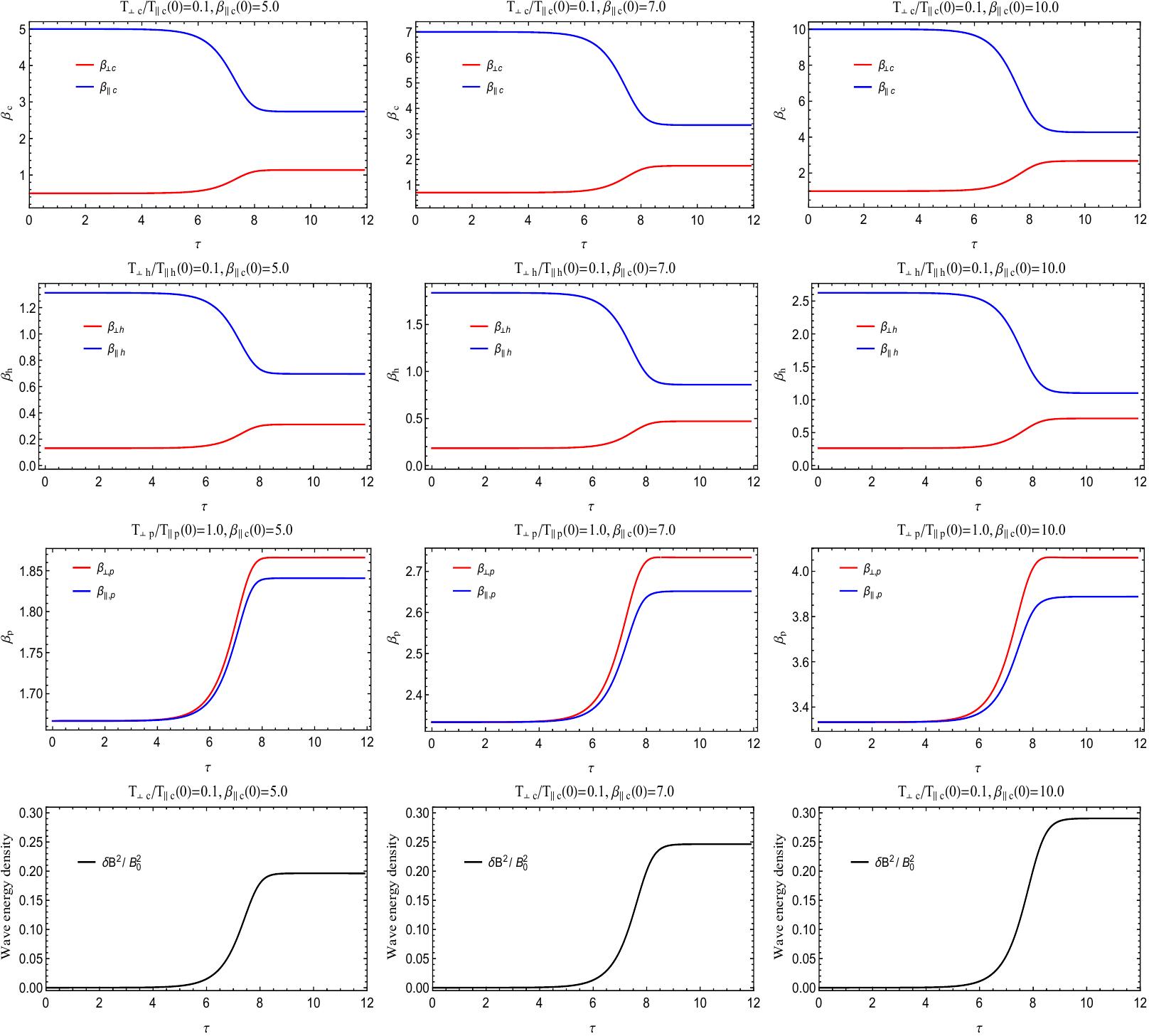}
\caption{Case 1: Time evolution of the normalized parallel and perpendicular plasma betas for 
the core electrons (top panels), halo electrons (middle-first), protons (middle-second), and the wave 
energy density (bottom panels) for different initial conditions: $\beta_{c,\parallel}(0) = 5.0, 7.0, 10.0$.}\label{f4}
\end{figure*}
%

\subsection{Quasilinear analysis}\label{sec:3.2}
%

Beyond the linear theory, the quasilinear (QL) analysis describes the temporal evolution of 
electromagnetic fluctuations, which are enhanced by the instability and turn to change the main
features of the distributions, such as temperature components ($T_{\perp,\parallel}$), the 
anisotropy ($A_a=T_\perp/T_\parallel$), and implicitly the plasma betas ($\beta_\perp/
\beta_\parallel\equiv T_\perp/T_\parallel$).However, we assume no exchange of electrons 
between core and halo, i.e. $\eta_h/\eta_c$ is constant, and the shape of the distributions
preserves, for instance, the protons and electron core remain Maxwellian, while the electron 
halo remain Kappa-distributed with the same value of power-index. The recent comparisons of the QL theory with the particle-in-cell simulation by \cite{Yoon2017a} and \cite{Lazar2018a}, show a reasonable agreement for the temporal evolution of the electron VDFs. There are also some indications from Vlasov simulations that power-index does not change much in the QL relaxation, though these  results are still limited to studies of different instabilities \citep{Lazar2017vlasov}. We resolve the system of QL equations (\ref{e11}) and (\ref{e12}) for the same plasma parameters introduced here above as cases 1 and 2.

%
\begin{figure*}[t]
\centering
\includegraphics[scale=0.78, trim={4.1cm 6.2cm 2.5cm 6.9cm},clip]{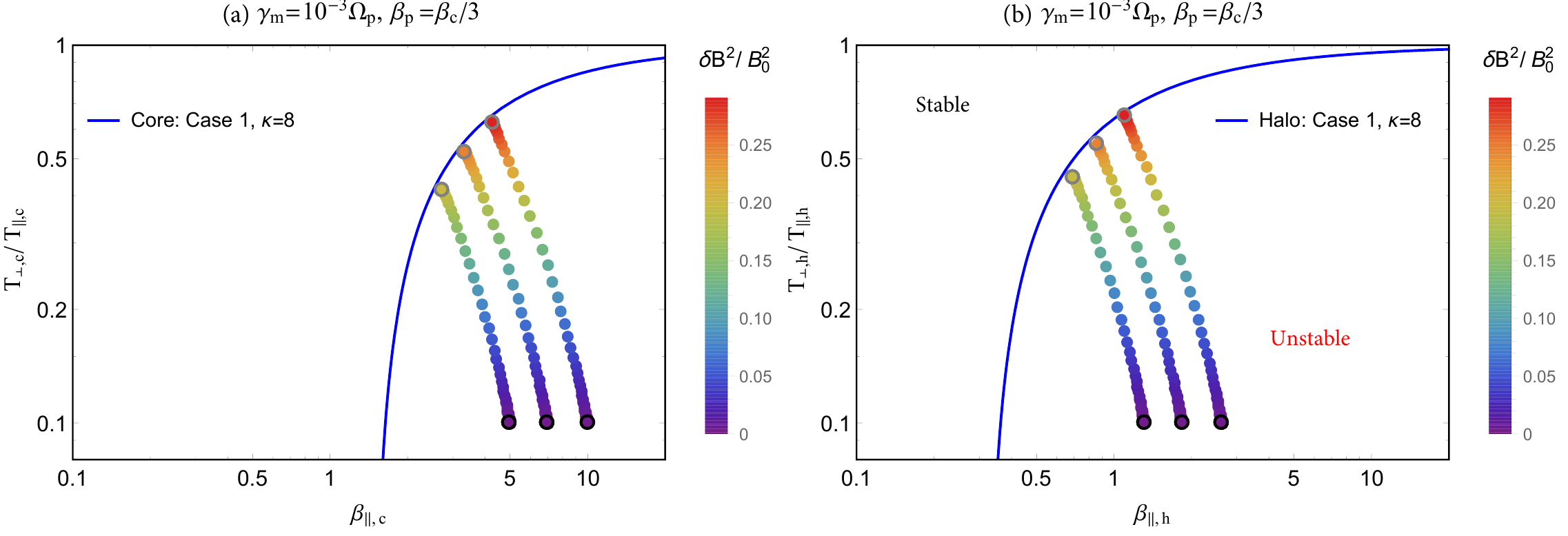}
\caption{Case 1: Dynamical paths for the core (panel a) and
halo (panel b) anisotropies showing the level of magnetic field fluctuations with color levels.
Initial states are indicated by black circles, while the final positions are marked with gray 
circles.}\label{f5}
\end{figure*}
%

Figures~\ref{f4} describes the time evolution of the perpendicular (red) and parallel (blue) plasma 
betas for the electron components, core (top) and halo (middle-first), thermal protons (middle-second),
and the corresponding variation of the wave energy density $\delta B^2/ B_0^2$ (bottom) as 
functions of the normalized time $\tau=\Omega_p~t$ for case 1 ($\kappa=8$) and different initial 
plasma betas $\beta_c(0)=5.0$ (left), 7.0 (middle), and 10.0 (right). 
The excitation of the EFH instability from the interplay of the anisotropic core and halo can 
regulate the initial temperature anisotropy for both these two components and also for protons,  
through the cooling and heating mechanisms reflected by the parallel (blue) and perpendicular 
(red) plasma betas. The level of saturated fluctuations is increasing with increasing the 
initial plasma beta, i.e. $\beta_c(0)=10.0$, confirming a stimulation of the instability
predicted for the peaking growth rates by the linear theory in Figure \ref{f2}(a). 
Nevertheless, both the core and halo components
show similar intervals of relaxation, but the halo remains slightly less anisotropic than the core after the
instability saturation, i.e., $\tau_m=12$. Linear theory cannot describe quantitatively the energy 
transfer between plasma particles mediated by the instability. However, from a quasilinear analysis 
we can estimate these exchanges induced by the free energy of electrons, which is transferred 
during the instability growing to the protons: Third row of panels shows the time evolution of 
the parallel and perpendicular plasma betas for protons ($\beta_{\perp, \parallel, p}$), which are initially
isotropic, i.e., $A_p(0)=1$. Both beta components are increased as $\tau$ increases, but protons
are heated more in perpendicular direction and become anisotropic at later stages $A_{p}(\tau_m)>1.0$. 
Their anisotropy increases with increasing the intial plasma beta.
These evolutions confirm, see \citep{Paesold1999, Messmer2002, Paesold2003}, that a finite amount of 
free energy is transferred by the growing fluctuations to the protons, especially in direction 
perpendicular to the background magnetic field. 

Figure~\ref{f5} presents temporal profiles of the temperature anisotropy and parallel plasma betas 
in a $(A_a, \beta_{\parallel, a})-$space, for both the core (subscript "$a=c$", in panel a) 
and halo (subscript "$a=h$", in panel b). Black circles indicate the initial positions, while the gray 
circles mark the limit states after saturation. The associated wave energy density $\delta B^2/ B_0^2$ 
is coded with colors. As expected, final states align along the anisotropy thresholds predicted by 
the linear theory, with unstable regimes located below the thresholds. This time evolutions clearly
shows deviations from isotropy decrease towards less unstable regimes closer marginal stability. These 
relaxations of the core and halo anisotropies are a direct consequence of the EFH instability 
and the enhanced electromagnetic fluctuations, which scatter the particles towards quasi-stationary 
states. This is confirmed by the increase of the wave energy density as the electrons become less anisotropic.      
%
\begin{figure*}[t]
\centering
\includegraphics[scale=1.1, trim={2.5cm 12.3cm 2.cm 2.6cm},clip]{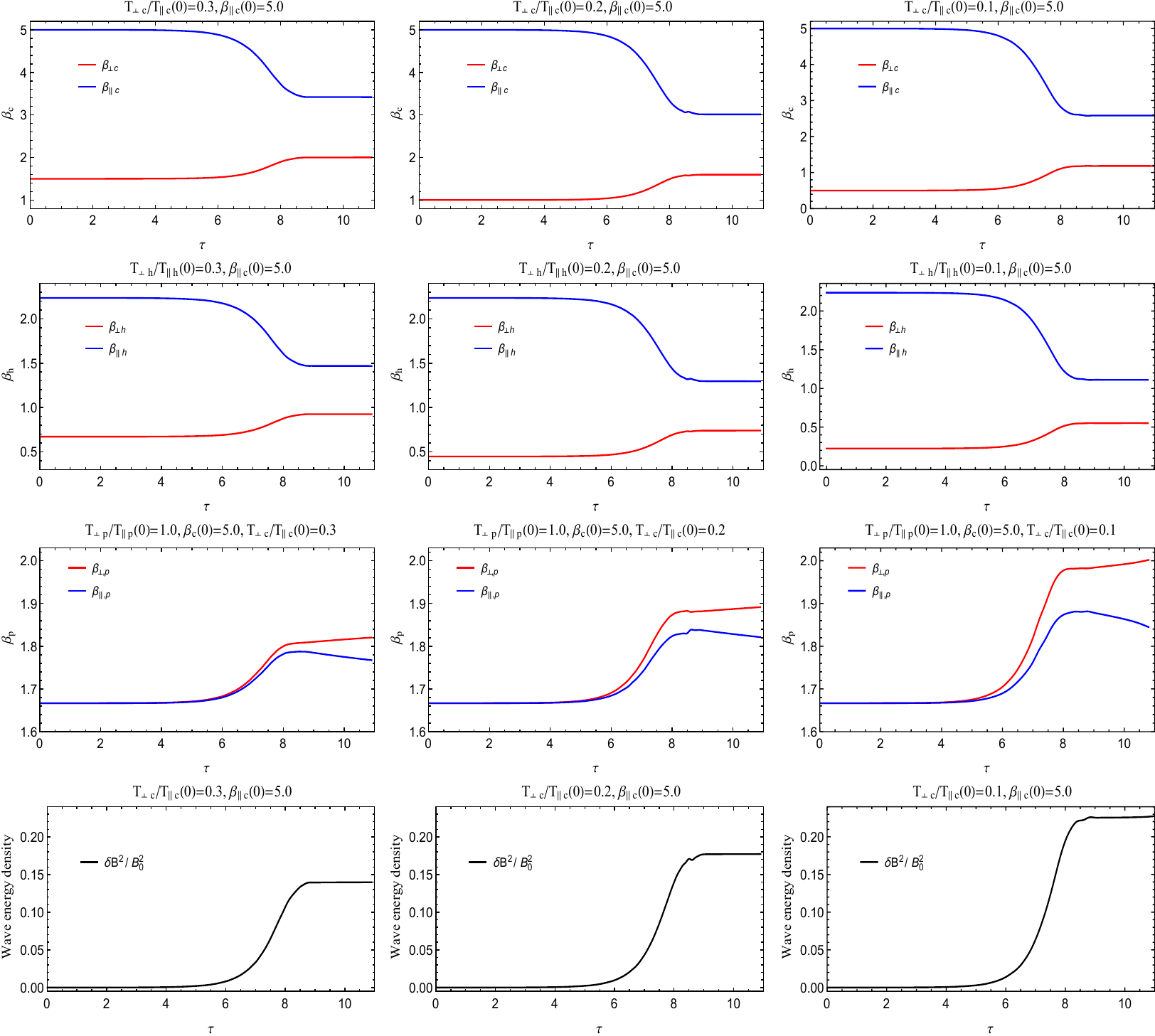}
\caption{Case 2: Time evolution of the normalized parallel and perpendicular plasma betas for 
the core electrons (top), halo electrons (middle-first), protons (middle-second), and the wave 
energy density (bottom) for different initial conditions given explicitly in the panels.} \label{f6}
\end{figure*}
%
%
\begin{figure*}[t]
\centering
\includegraphics[scale=0.755, trim={4.1cm 6.2cm 2.5cm 6.9cm},clip]{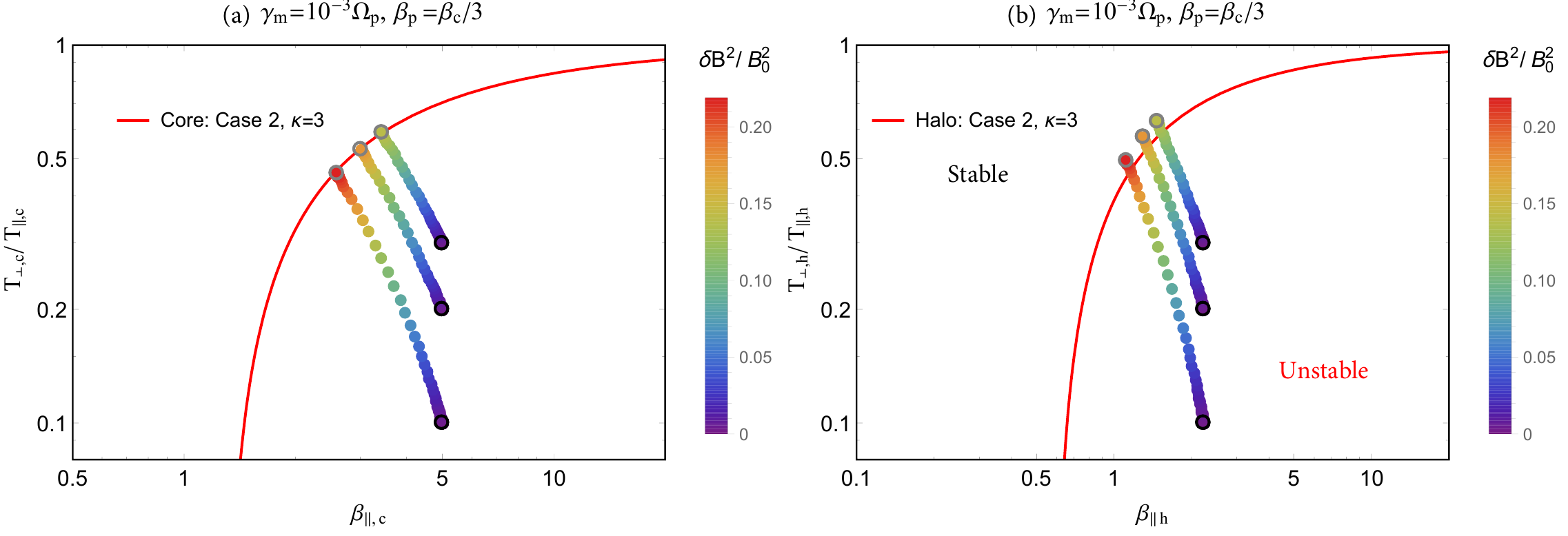}
\caption{The same as in Figure.\ref{f5} but for Case 2.}\label{f7}
\end{figure*}
%

%
\begin{figure*}[t]
\centering
\includegraphics[scale=1.04, trim={2.6cm 15.8cm 2.cm 2.7cm},clip]{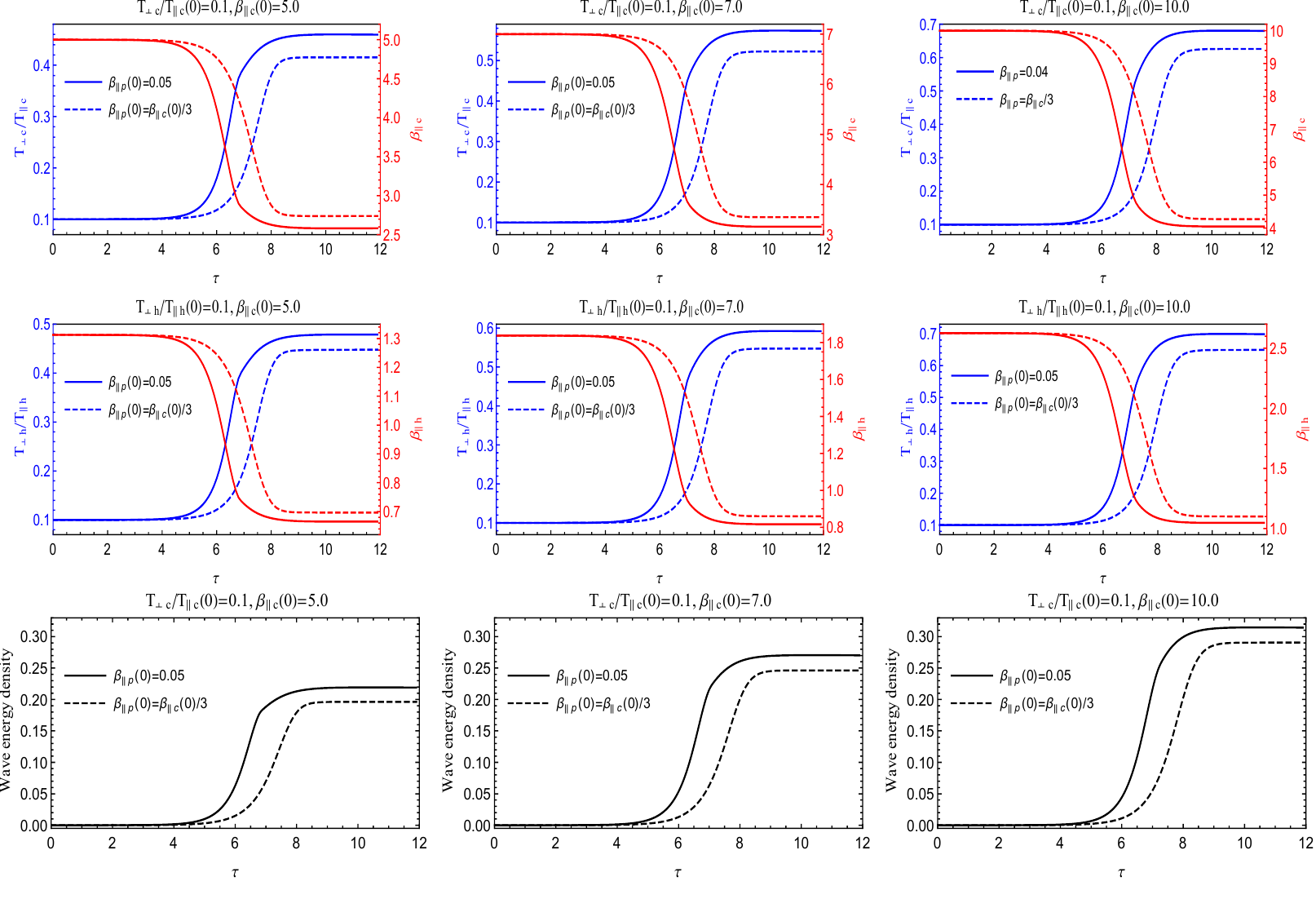}
\caption{Comparison between the effects of protons in Case 1 with $\beta_{\parallel p}(0)=
\beta_{\parallel c}(0)/3$ (dashed lines) and protons with relatively low plasma beta 
$\beta_{\parallel p}(0)=0.05$ (solid lines).}\label{f8} 
\end{figure*}
%

The results in Figure~\ref{f6} are obtained for case 2 ($\kappa=3$), 
with an initial $\beta_c(0)=5.0$ and different anisotropies $A_c(0)=A_h(0)=0.3$ (left), $0.2$ (middle), $0.1$ (right). The 
main effects, associated with the heating or cooling of different particle populations by the 
enhanced fluctuations, are reflected by the temporal profiles of the beta components for core, 
halo, and protons. These effects are similar to those in case 1, but here are mainly triggered 
by the different anisotropies, e.g., for higher anisotropies,  i.e.\ $A_{c,h}(0) =~0.1$ (right column), the resulting magnetic wave energy $\delta B^2/ B_0^2$ increases as an effect of the enhancement of the instability growth rates predicted by the linear theory in Figure~\ref{f2b} (a). This may also explain the increase of the energy transferred to the initially isotropic protons ($A_p(0)=~1.0$), which reach moderate anisotropies $A_p(\tau_m)>1.0$ after saturation.

In Figure~\ref{f7} we display dynamical paths of the temperature anisotropy and the parallel plasma beta 
in a $(A_a, \beta_{\parallel, a})-$space for both the core (panel a) and halo (panel b). The initial 
anisotropies $A_a(0)$ are marked with black circles, while anisotropies $A_a(\tau_m)$ at final stages 
are indicated by gray circles. The levels of the magnetic wave energy $\delta B^2/ B_0^2$ are coded 
with colors. As in case 1 (Figure~\ref{f5}), the initial deviations from isotropy are reduced in time,
towards less unstable states. It is obvious that final states tend to align to the instability thresholds,
and larger initial anisotropies, e.g., $A_a(0)=0.1$, will determine longer dynamical path for the particles. Nevertheless,  in Figure~\ref{f7} the final states of electron anisotropies reach a more stable regime by comparison to  Figure~\ref{f5}. One plausible explanation for this physical behavior can be found in the increase of the suprathermal population which stimulate the energy exchange during the instability development to protons, enhancing their anisotropy ($T_\perp > T_\parallel$), while  the core and halo electrons become less anisotropic. This explanation is also supported by the results in Figure~\ref{f4}, left panels, and Figure \ref{f6}, right panels, which are obtained for $\beta_c(0)=5.0$ and  $A_e(0)=0.1$, but different values of the power-index  $\kappa=8$  ( Figure~\ref{f4}) and $3$ (Figure \ref{f6}). It is obvious that at later stages both the  proton anisotropy and the  associated magnetic wave energy obtained for $\kappa=3$. i.e.\ $\delta B^2/ B_0^2=0.24$ and $A_p(\tau_m)=1.08$, are higher than those obtained for $\kappa=8$, i.e.\ $\delta B^2/ B_0^2=0.2$ and $A_p(\tau_m)=1.016$.

\begin{figure}[t]
\centering
\includegraphics[scale=0.45]{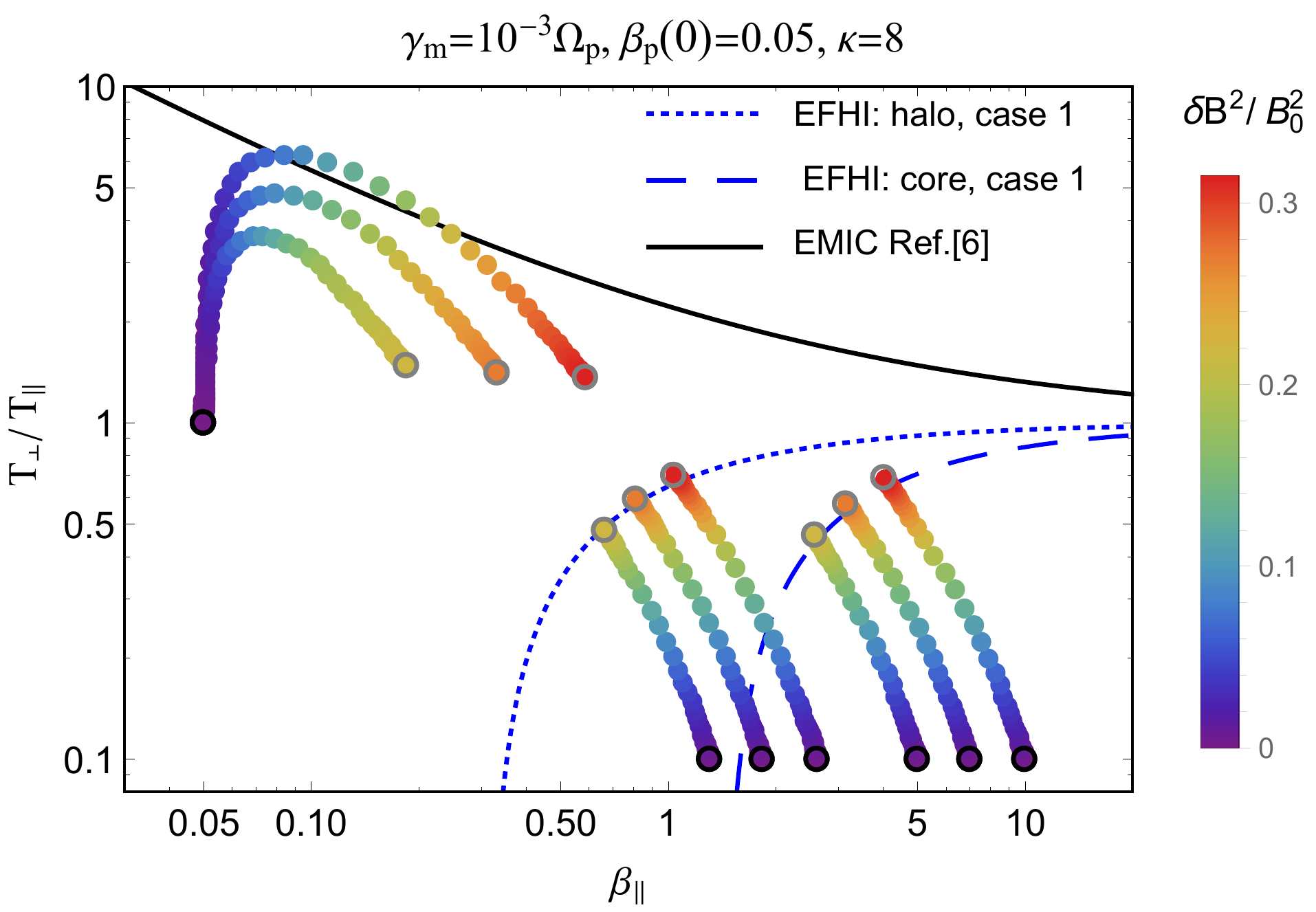}
\caption{Dynamical paths of case 1, but with $\beta_{\parallel, p}=0.05$.}\label{f9}
\end{figure}

In both these two cases the initial proton plasma beta $\beta_{\parallel p}(0)~=\beta_{\parallel c}(0)/3>1.0$
is relatively large, and transfer of energy from electrons to protons remains modest. Dynamical paths of 
the proton anisotropy are very short in these cases, and we have not displayed them in Figures~\ref{f5} and 
\ref{f7}. Seeking completeness, we extend the analysis for a series of new cases in Figures \ref{f8} and 
\ref{f9}, which allow us to examine the mutual effects between the electron populations and protons with 
relatively low initial $\beta_{\parallel p}(0)=0.05$. 

Figure~\ref{f8} presents by comparison temporal profiles of temperature anisotropy for the core (top) and 
halo (middle), as well as the associated magnetic wave energy (bottom) for the same plasma parameters 
in case 1, but for two different initial proton plasma betas, either large $\beta_{\parallel, p}(0)=
\beta_{\parallel c}(0)/3>1.0$ (dotted lines), or small $\beta_{\parallel p}(0)=0.05$ (solid lines). 
It is clear that protons with lower beta stimulate the cooling process of both populations of electrons, 
making it faster, and the core and halo electrons become less anisotropic in this case (solid blue lines). 
The associated wave energy is enhanced with decreasing proton parallel beta, confirming the enhanced 
relaxation of the anisotropies. The dynamical paths of the core, halo and proton anisotropies for the new 
case with lower $\beta_{\parallel p}(0)=0.05$ are displayed in Figure~\ref{f9}. For the core and halo 
the paths are orientated  in the direction of stable regimes and end up or close to the anisotropy 
thresholds (blue lines) predicted by the linear theory in Figure \ref{f3}. This behavior suggests that 
for low plasma betas $\beta_{\parallel,p}=0.05$, the free energy of electrons is transferred via the 
growing fluctuations to the (resonant) protons, and especially in perpendicular direction. Thus, at 
later stages after saturation both core and halo become less anisotropic, while the initially isotropic 
protons become strongly anisotropic $A_p(\tau_m)>1$, which is evident in Figure~\ref{f9}. These 
variations are stimulated by increasing  the initial plasma beta, i.e.\ $\beta_{\parallel c}(0)=10.0$. 
Protons with temperature anisotropy $A_p>1$ may trigger LH polarized electromagnetic ion cyclotron 
(EMIC) instability with a maximum growth rate in the parallel direction to the background magnetic 
field \citep{Shaaban2015, Shaaban2016}. \cite{Shaaban2017} show that an anisotropic electrons with 
$A_e<1.0$  and their suprathermal populations may stimulate the EMIC unstable modes. These 
predictions from linear theory seems to be confirmed and explained by the present quasilinear analysis. 
For a visual guidance Figure~\ref{f9} displays also the EMIC anisotropy threshold 
(black line) derived by \cite{Shaaban2017}, with fitting parameters $s=-1.221$ and $\alpha=0.579$ 
in Eq.(\ref{e17}). For $\beta_{\parallel c}(0)=0.05$, protons (with $A_p(0)=1$) gain anisotropy in 
perpendicular direction and move towards the EMIC thresholds, and then their anisotropy decreases to 
settle down much below the instability threshold after saturation.

From a comparison of our results in Figure \ref{f9} with those in a recent study by 
\citet[Figure 4 therein]{Sarfraz2018}, we find a reasonable agreements for the dynamical paths of 
the proton anisotropies, but not for the instability thresholds and paths predicted 
for the core and halo electrons by the linear and quasilinear approaches.  
Contrary to \cite{Sarfraz2018} the halo anisotropy threshold is lower than that obtained for the 
core anisotropy threshold, and extends to lower parallel plasma beta. Our results, from both 
linear and the quasilinear analyses, show an excellent agreement with the observational limits of 
the core and halo anisotropies, and this mainly explains by the more realistic plasma parametrization 
used in our analysis.

%

\section{Conclusions}\label{sec:4}

In-situ measurements of the velocity distributions of solar wind electrons reveal two central 
components, a bi-Maxwellian thermal core and a bi-Kappa suprathermal halo \citep{Stverak2008}. 
Recently \cite{Pierrard2016} have also shown that the core and halo electrons are not independent 
of each other, and their properties may show direct correlations. These correlations allow us
to characterize the instability in terms of either the core or halo parameters. Of particular
interest are the highest growing modes which arise when both the core and halo electrons 
exhibit similar anisotropies, in our case $A_{a=h,c}=T_{a,\perp}/T_{a,\parallel}<1$. The firehose 
instability conditions are significantly changed by the interplay of these two populations, such 
that the thresholds of the periodic branch approach very well the limits of temperature anisotropy 
reported by the observations for both the core and halo electrons. 
 
We have performed a numerical analysis for both linear dispersion and quasilinear equations 
derived in Section~\ref{sec:3}, for conditions typically encountered in the slow wind. 
Mutual effects of the core and suprathermal halo can markedly stimulate the instability and 
induce new unstable regimes instability, lowering the instability thresholds by comparison 
to those provided by the idealized theories of Maxwellian distributed plasmas. The same effects 
are reflected by the quasilinear saturation of the instability and the relaxation of temperature 
anisotropy towards the anisotropy thresholds. The evolution of both the core and halo electrons is 
characterized by perpendicular heating and parallel cooling, lowering their anisotropies 
during the development of the firehose instability. The relaxation of the temperature anisotropy 
is found to be associated with an enhancement of the magnetic wave power before reaching the 
saturation. Protons gain energy from the left-handed electromagnetic fluctuations, especially 
in direction transverse to the magnetic field, and this energy transfer becomes more important
for protons with a low initial plasma beta (i.e., $\beta_p(0)=0.05$). 

In this study, we have restricted ourselves to the unstable periodic firehose modes that develop with a highest growth rate in directions parallel to the background magnetic field.  A QL approach of the aperiodic firehose modes growing faster in the oblique directions \citep{Li2000, Gary2003a, Shaaban2018fh} is not yet feasible, but our present results will certainly stimulate an extended analysis. Furthermore, the influence of the solar wind inhomogeneity (e.g., \cite{Yoon2016a, Yoon2017b}) has not been considered in the present analysis, and their interplay with the effects of suprathermal electrons will be investigated in the future.

\begin{table}[h]
\caption{Fitting parameters for thresholds $\gamma_{\rm m}/\Omega_p = 10^{-2}$}\label{t1}
\centering 
\begin{tabular}[t]{ccc|cc|ccc}
\hline
$\kappa$ & \multicolumn{2}{c}{Core} & \multicolumn{2}{c}{Halo} & \multicolumn{2}{c}{SBM} \\ \hline
         & $s$  & $\alpha$ & $s$ & $\alpha$ & $s$   & $\alpha$ \\
$3$      & 1.32 &  0.994   & 0.6 &   1.03   &  --    & --  \\
$8$      & 1.45 &   1.0    & 0.35&   0.88   &  --    & -- \\
$\infty$ &  --  &   --     &  -- &   --     & 1.75  & 1.04 \\
\hline
\end{tabular}
\end{table}

\acknowledgments
The authors acknowledge support from the Katholieke Universiteit Leuven, Ruhr-University Bochum, and 
Alexander von Humboldt Foundation. These results were obtained in the framework of the projects 
SCHL 201/35-1 (DFG-German Research Foundation), GOA/2015-014 (KU Leuven), G0A2316N (FWO-Vlaanderen), 
and C 90347 (ESA Prodex 9). S.M. Shaaban would like to acknowledge the support by a Postdoctoral 
Fellowship (Grant No. 12Z6218N) of the Research Foundation Flanders (FWO-Belgium). P.H.Y. acknowledges the BK21 Plus grant (from NRF, Korea) to Kyung Hee University, and financial support from GFT Charity Inc., to the University of Maryland. Thanks are
due to {\v S}. {{\v S}tver{\'a}k} for providing the observational data.




\bibliography{papers}


%
\listofchanges
\end{document}